\documentclass[a4paper,12pt]{scrartcl}
\usepackage[utf8]{inputenc}
\usepackage[T1]{fontenc}
\usepackage[english]{babel}
\usepackage{csquotes}

\usepackage[section]{placeins}
\usepackage{graphicx}
\usepackage{caption}
\usepackage{subcaption}
\usepackage{varioref}
\usepackage[breaklinks=true,colorlinks=false, pdfborder={0 0 0}]{hyperref}

\usepackage{siunitx}

\usepackage[shortcuts]{extdash}

\usepackage{amsmath}
\usepackage{amsfonts}
\usepackage{amssymb}

\usepackage[backend=biber,style=numeric,autopunct, sorting=none, alldates=year, block=ragged, autolang=hyphen, isbn=false]{biblatex}
\usepackage[nottoc]{tocbibind}

\graphicspath{ {../Bilder/} }
\DeclareGraphicsExtensions{.pdf,.png,.jpg}

\title{ROPPERI \\ \large{ Readout Of a Pad Plane with ElectRonics designed for pIxels }\\-\\ \LARGE{ A TPC readout with GEMs, pads and Timepix }}
\author{Ulrich Einhaus\thanks{DESY Hamburg}, Jochen Kaminksi\thanks{University of Bonn}, Michele Caselle\thanks{KIT Karlsruhe}}
\date{\large Talk presented at the International Workshop on Future Linear Colliders (LCWS2017), Strasbourg, France, 23-27 October 2017. C17-10-23.2.}

\begin{document}

\maketitle

\begin{abstract}
The concept of a hybrid readout of a time projection chamber is presented. It combines a GEM-based amplification and a pad-based anode plane with a pixel chip as readout electronics. This way, a high granularity enabling to identify electron clusters from the primary ionisation is achieved as well as flexibility and large anode coverage.
The benefits of this high granularity, in particular for dE/dx measurements, are outlined and results of a simulation-based performance study are given.
The structure of the first prototype board is discussed, including adaptions based on a very preliminary first measurement for a second production towards a proof-of-principle.
\end{abstract}

\section{Introduction}
\label{sec:introduction}

\begin{figure}[thp]
  \begin{subfigure}{.48\textwidth}
    \centering
    \includegraphics[width=\textwidth,height=0.35\textheight,keepaspectratio=true]{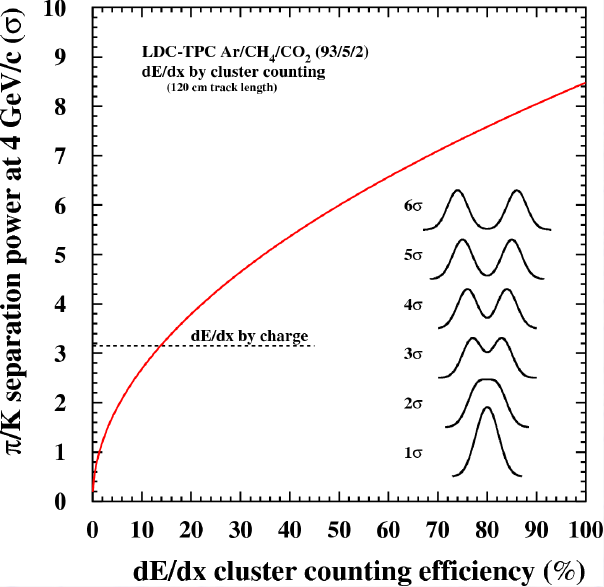}
  \end{subfigure}
  \centering
  \begin{subfigure}{.48\textwidth}
    \centering
    \includegraphics[width=\textwidth,height=0.35\textheight,keepaspectratio=true]{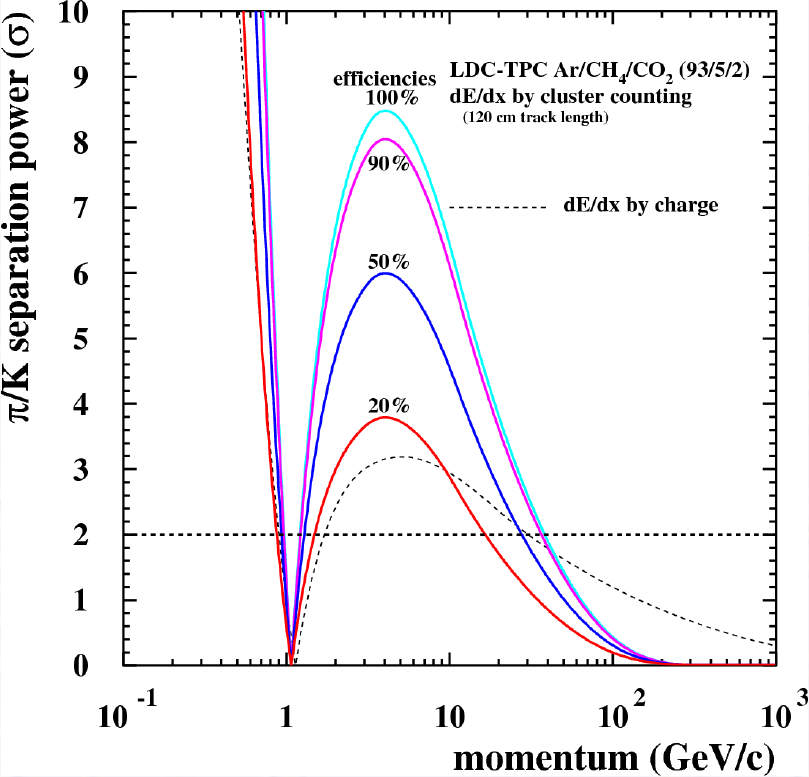}
  \end{subfigure}
  \caption{Pion/kaon separation power of dE/dx by cluster counting, from \cite{Hauschild06}.}
  \label{fig:piokaonsep}
\end{figure}

For the International Large Detector (ILD) \cite{ILD_TDR_detectors} at the International Linear Collider (ILC) \cite{ILCSummary} a Time Projection Chamber (TPC) \cite{TPC_CC} is foreseen as central tracker. A key feature of this gaseous detector is the inherent dE/dx capability: Since the relation of momentum and energy loss (dE/dx) of a traversing particle depends on its rest mass, and thus its species, measuring both properties allows for a particle identification determined by the Bethe-Bloch equation. The energy loss is conventionally measured by summing all electrons generated from the ionisation by the incident particle. For each ionising interaction the number of generated electrons is given by a Landau distribution which has a long tail towards large numbers of electrons. The relatively large width of this distribution worsens the correlation of the measured energy and the momentum of the particle. It is advantageous to instead count the number of ionising interactions of the incident particle. This is given by a Poissonian distribution with a significantly smaller width, resulting in a better correlation and particle identification power, as demonstrated in \cite{Hauschild06}. In \autoref{fig:piokaonsep}, the separation power for pion/kaon-separation depending on the cluster counting efficiency is shown compared to the conventional dE/dx by charge summation. In former experiments with prototypes, a cluster counting efficiency of only \SI{20}{\percent}-\SI{30}{\percent} was reached. Nevertheless, the resulting separation power is still better than by charge summation. Also, improved algorithms are expected to deliver a higher cluster counting efficiency. However, cluster counting can only work if a sufficient correlation between the position of the electrons of one cluster is preserved during drift. This is investigated in simulation in \autoref{sec:simulation}.

\begin{figure}[thp]
   \centering
   \includegraphics[width=0.95\textwidth,height=0.3\textheight,scale=1,keepaspectratio=true]{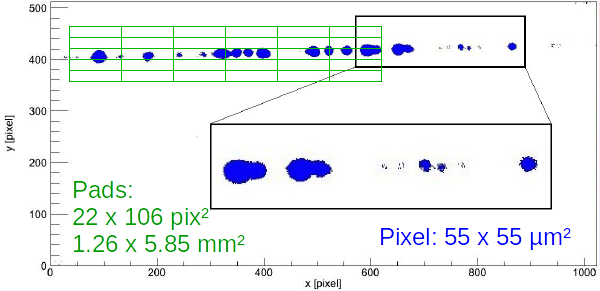}
   %Event display - GEM+Timepix+Pad grid+Labels.png: 378x340 pixel, 72dpi, 13.34x11.99 cm, bb=0 0 378 340
   \caption{Event display of a track recorded with a Timepix Octoboard, from \cite{Lupberger14:1}. The activated pixels are shown in blue, the green overlay shows the pitch of a typical pad-based readout.}
   \label{fig:event_display}
 \end{figure}

To achieve the cluster counting capability, a sufficiently high granularity is needed for the TPC readout system. A comparison of two existing prototype systems is shown in \autoref{fig:event_display}. The electron clusters are amplified by GEMs (Gas Electron Multipliers) \cite{SAULI1997531}, creating charge clouds visible as blue blobs. The anode consists of 8 Timepix ASICs with their pixels as immediate sensitive anode, which has a pitch of \SI{55x55}{} \si{\micro\meter ^2}. The charge clouds are clearly identifiable – the granularity is even higher than needed, causing more data than required. The green overlay of a typical current pad-based readout system (GridGEM \cite{lctpc16}), shows, that with its pads of ca. \SI{1x6}{} \si{mm^2} a cluster identification is however not possible. Therefore, we propose a novel readout structure, called ROPPERI (Readout Of a Pad Plane with ElectRonics designed for pIxels), that allows pad sizes of around \SI{300}{\micro\meter} to enable cluster counting, gives a large flexibility and keeps the channel number low at the same time.

\section{Readout Board Structure}
\label{sec:board}

\begin{figure}[thp]
  \centering
  \includegraphics[width=0.95\textwidth,height=0.25\textheight,scale=1,keepaspectratio=true]{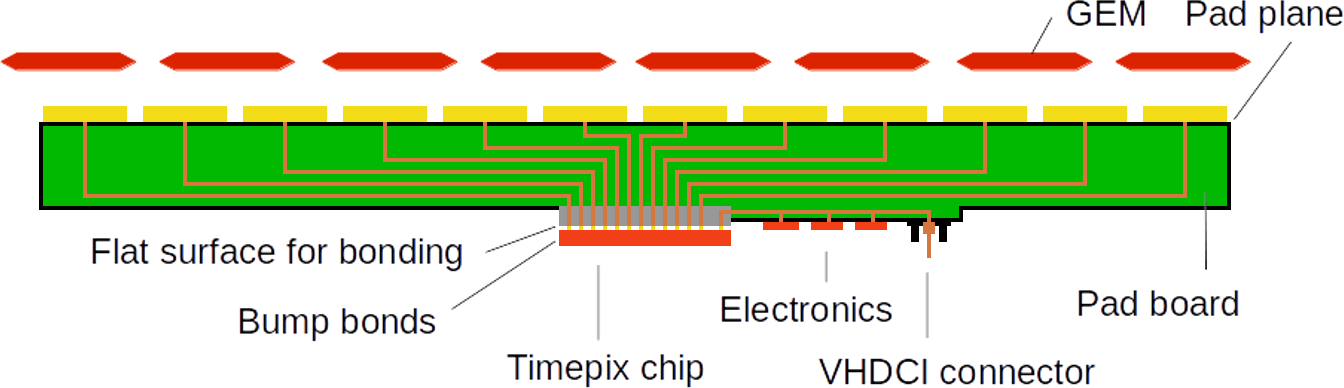}
  %Event display - GEM+Timepix+Pad grid+Labels.png: 378x340 pixel, 72dpi, 13.34x11.99 cm, bb=0 0 378 340
  \caption{ROPPERI schematics.}
  \label{fig:routing_structure}
\end{figure}

\begin{figure}[thp]
  \centering
  \includegraphics[width=0.95\textwidth,height=0.4\textheight,scale=1,keepaspectratio=true]{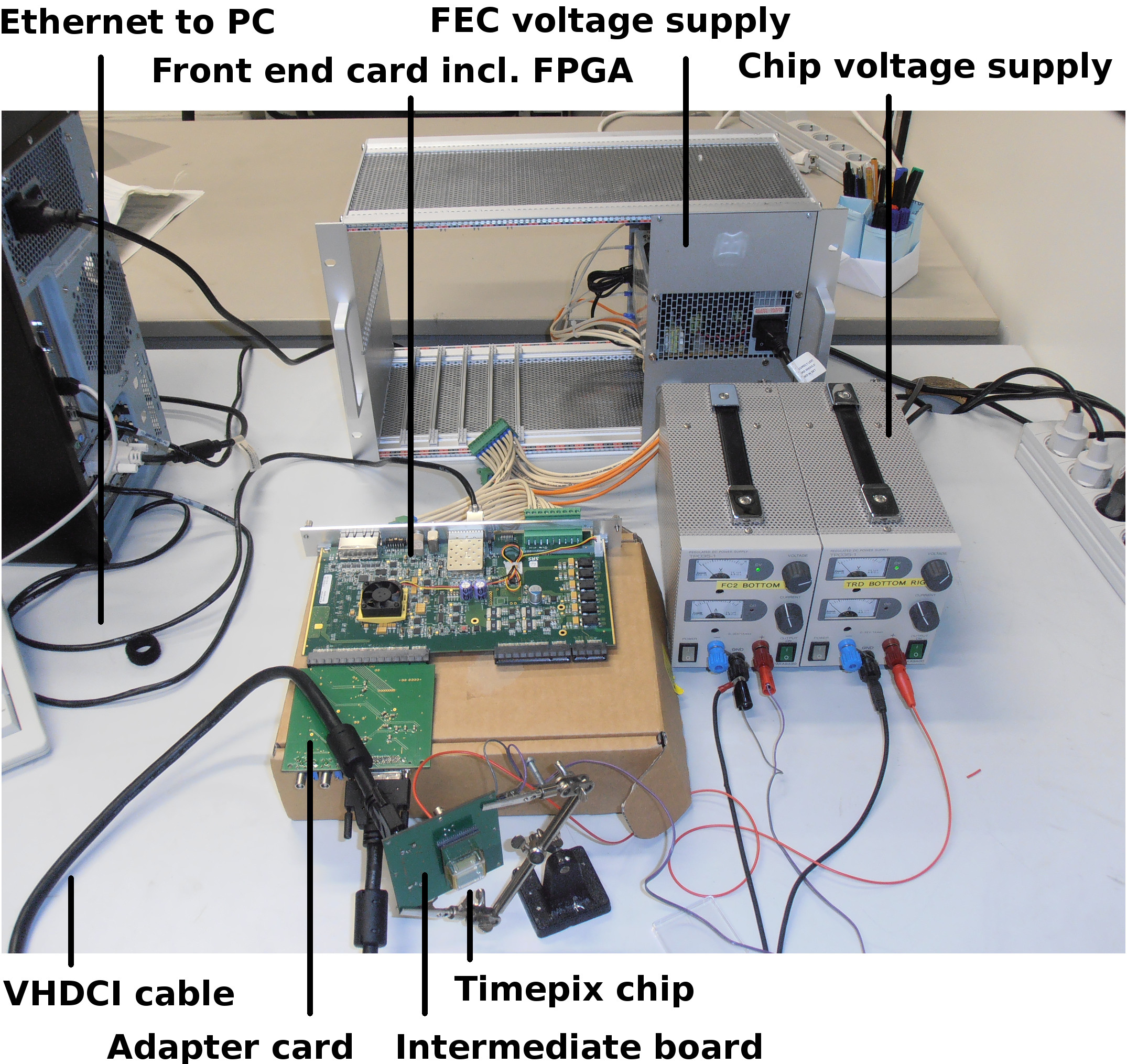}
  %Event display - GEM+Timepix+Pad grid+Labels.png: 378x340 pixel, 72dpi, 13.34x11.99 cm, bb=0 0 378 340
  \caption{Timepix readout setup.}
  \label{fig:full_setup}
\end{figure}

The ansatz for a ROPPERI board implementation is shown in \autoref{fig:routing_structure}: GEMs are used for amplification, small pads on a PCB form the anode and are read out by a Timepix ASIC, which has a matrix of \SI{256x256}{} $=65,536$ pixels. The connections from the pads are routed through the PCB to the ASIC which is bump bonded to the PCB surface, that needs to be sufficiently flat. Technology-wise, it is pioneering work to connect a pixel chip with such a small pitch directly to a PCB. The Timepix power and communication pads are on the same side of the ASIC as the pixels. They are usually connected by wire bonds. In the ROPPERI approach also these pads have to be connected by bump bonds to the PCB, which in addition hosts the further electronics elements including the connectors for the chip voltage supply and an I/O-cable plug. The data processing is conducted by the SRS (Scalable Readout System) developed by the RD51 group at CERN \cite{SRS_RD51_11}. A Front-End Concentrator card (FEC) hosts a Field Programmable Gate Array (FPGA) that reads the data from the Timepix ASIC via an adapter card and a VHDCI-cable to the ROPPERI board, for a detailed implementation see \cite{Lupberger16}. The FEC can be connected via Ethernet to a PC, see \autoref{fig:full_setup}.

\subsection{Benefits Compared to Existing Systems}
\label{sec:benefits}

Compared to the pad-based systems, like the GridGEM \cite{lctpc16}, ROPPERI can have a higher granularity leading to an improved double hit/track resolution, to a reduced occupancy and in particular to the charge cloud and possibly cluster identification capability. In addition, the Timepix ASIC allows for a significantly smaller footprint of the readout electronics.\\
Compared to the pixel-based systems, in particular the InGrid system \cite{Lupberger14:2}, ROPPERI has a lower granularity, but measures each GEM charge cloud with several pads, allowing for a position calculation using a fit. Instead, the InGrid system records each electron with exactly one pixel. Pixel granularity combined with a GEM, as in \cite{Renz09}, does not increase the performance, but only the data output. In addition, the pixel-based anode can currently only cover ca. \SI{50}{\percent} of the anode area in a setup with 12 octoboards \cite{Lupberger11}, or up to \SI{63}{\percent} with tighly stacked ASICs as suggested in \cite{Timmermans11}, since the ASICs need some overhead area, and the module geometry of the foreseen ILD TPC is not square, as the ASICs are. ROPPERI's separate anode PCB allows for a coverage of more than \SI{90}{\percent}, comparable to the traditional pad-based systems. Also, it is more flexible with regard to the granularity. For a desired change in pad size it is required to only produce new PCBs and bond them to the ASICs and not to produce new ASICs with a different pitch.

\section{Hardware}
\label{sec:hardware}

\subsection{Current Status}
\label{sec:hardware_status}

\begin{figure}[thp]
  \centering
  \includegraphics[width=0.95\textwidth,height=0.4\textheight,scale=1,keepaspectratio=true]{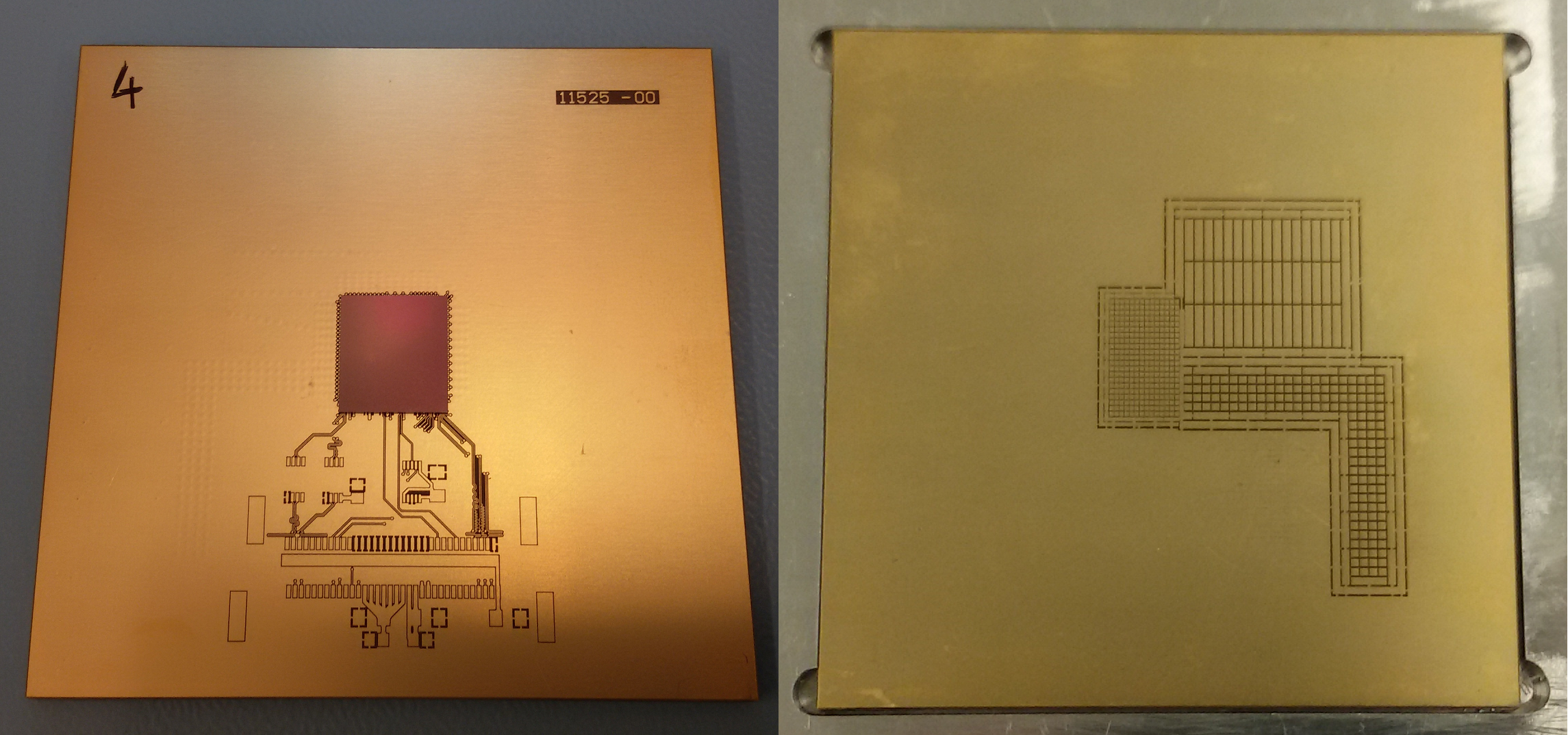}
  \caption{Foto of the first board. Left: Timepix bonded to the 'lower side' and solder pads for electronics elements. Right: sensitive pads on the 'upper side', pointing to the drift volume.}
  \label{fig:Board_both_sides}
\end{figure}

\begin{figure}[thp]
  \centering
  \includegraphics[width=\textwidth,height=0.75\textheight,keepaspectratio=true]{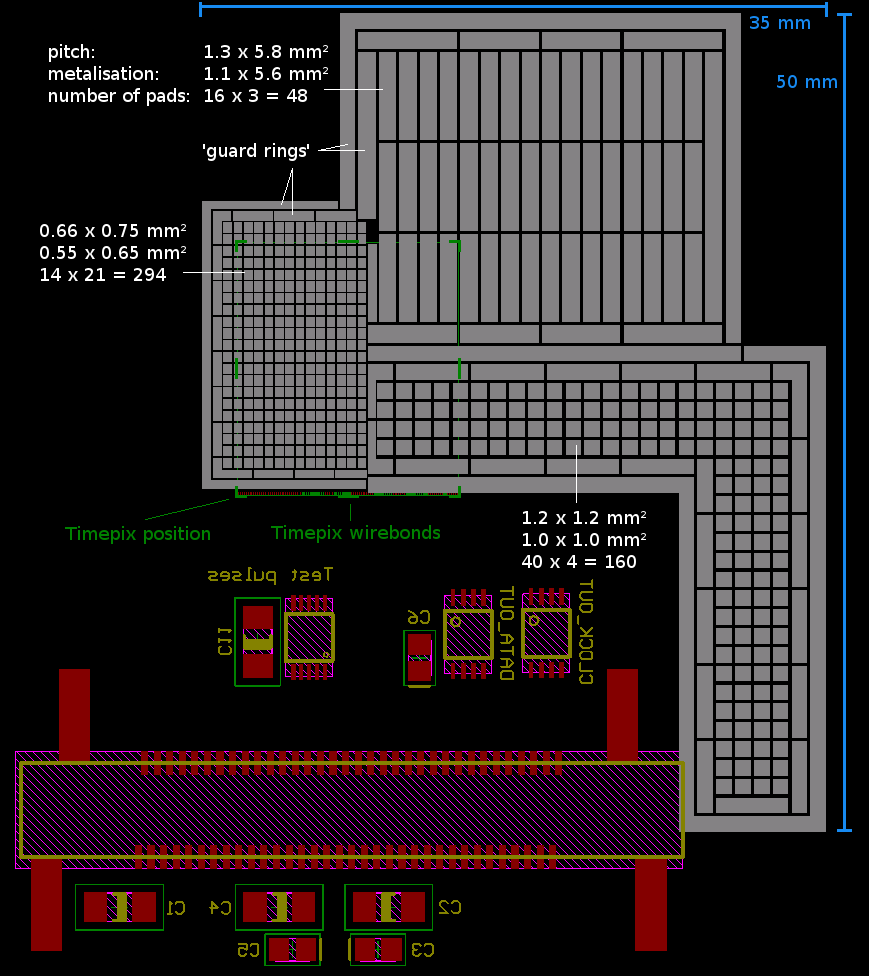}
  \caption{CAD layout of the first ROPPERI prototype board.}
  \label{fig:padlayout}
\end{figure}

%The SRS has been set up and successfully tested. For this, a single Timepix ASIC was mounted close to an electrically pulsed coin and read out. Through the capacitative coupling a signal was induced to the pixels. The resulting ADC data map of the chip shows the coin, see \autoref{fig:TP_cent}. The dead columns and noisy pixels are part of the low-quality chip used for the tests. For the actual test system, high quality chips are available with a significantly lower count of dead or noisy channels.
The first prototype board was designed, produced, equipped, bonded with a Timepix ASIC and tested. \autoref{fig:Board_both_sides} shows pictures of top and bottom sides of the board, in this case with a bonded chip but not equipped. The CAD layout in \autoref{fig:padlayout} shows the main features of the first board: Three different pad sizes are used for the sensitive pads, shown in grey. The currently minimal achievable pad pitch is \SI{0.66x0.75}{} \si{mm^2}, used for about 300 of the pads placed directly above the Timepix ASIC (envelope shown in green). The \SI{1.2x1.2}{} \si{mm^2} pads are used to check signal quality for different line lengths. An array of \SI{1.3x5.8}{} \si{mm^2} pads corresponds to the standard pad pitch of FADC systems for comparison. The sensitive pads are surrounded by grounded 'guard ring' pads to reduce cross talk. In total 502 pads are connected to the chip and cover a sensitive area of a few \si{cm^2}. In the lower part of \autoref{fig:padlayout} the placements of some passive elements and the VHDCI connector are shown in red. The board is going to be used with a triple GEM stack of standard CERN \SI{10x10}{} \si{cm^2} GEMs in a modular prototype TPC. This prototype has a maximum drift length of 5 \si{cm} and can be used with an integrated radioactive source.

\subsection{Bonding}
\label{sec:hardware_bonding}

\begin{figure}[thp]
  \centering
  \includegraphics[width=\textwidth,height=0.75\textheight,keepaspectratio=true]{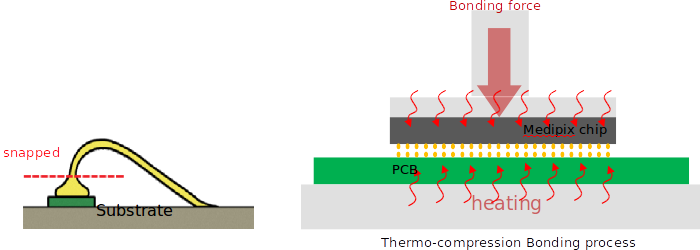}
  \caption{Schematic of bonding procedure. Left: applying a bump to a substrate; right: flip-chip process.}
  \label{fig:bonding_schematic}
\end{figure}

\begin{figure}[thp]
  \centering
  \includegraphics[width=\textwidth,height=0.75\textheight,keepaspectratio=true]{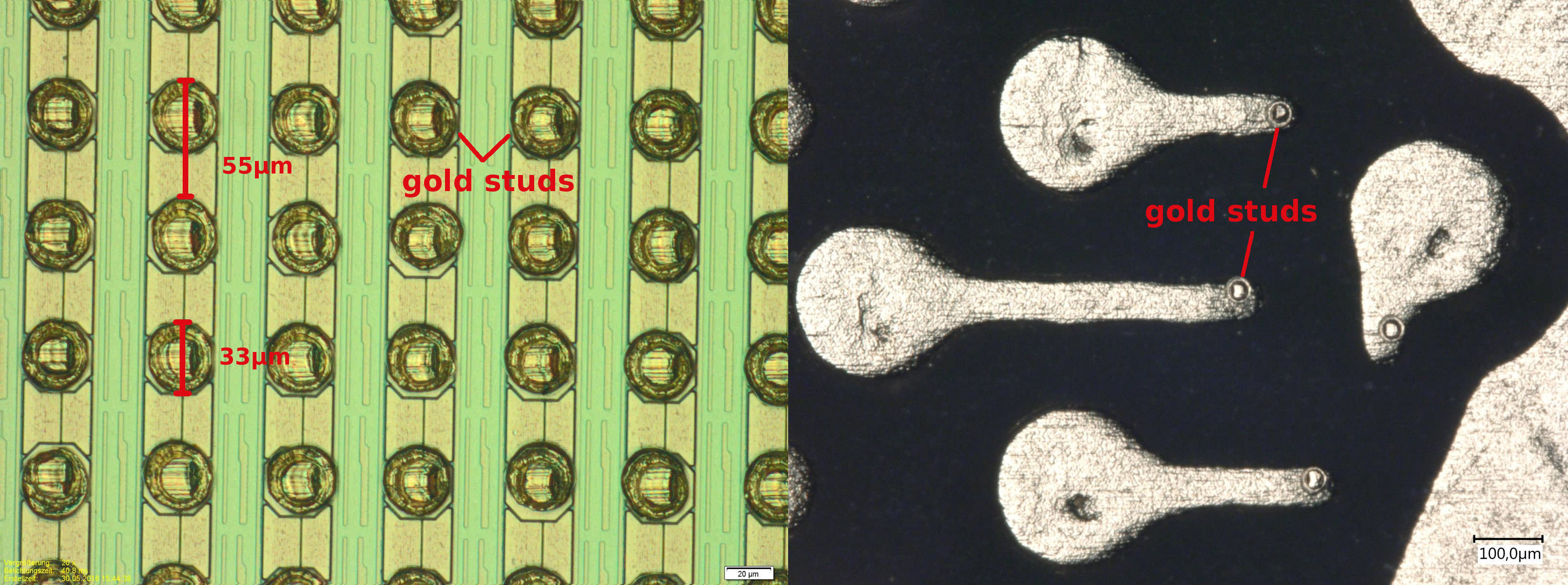}
  \caption{Result of bumping procedure. Left: gold studs on Timepix; right: gold studs on PCB.}
  \label{fig:bonding_substrates}
\end{figure}

\begin{figure}[thp]
  \centering
  \includegraphics[width=\textwidth,height=0.75\textheight,keepaspectratio=true]{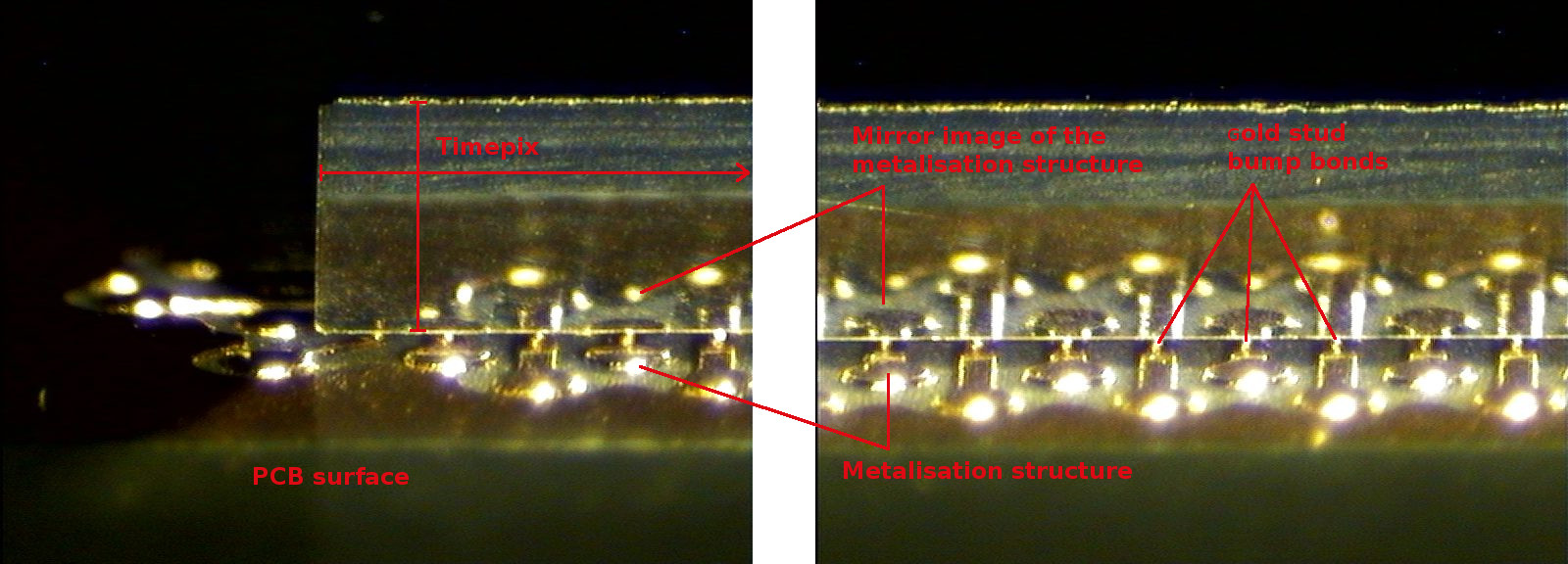}
  \caption{Result of bonding procedure: Bonding gap in a sideview.}
  \label{fig:bonding_sideview}
\end{figure}

The Timepix ASIC was bonded to the PCBs at the IPE bonding lab at the Karlsruhe Institute for Technology (KIT).
In the gold stud bump bonding process, a \SI{15}{\micro\meter} diameter gold wire is bumped to the PCB bond pads as well as the respective aluminium openings in the chip surface. A \SI{30}{\micro\meter} diameter bump, the gold stud, forms and the remaining wire is snapped off with a sharp edge, see \autoref{fig:bonding_schematic}, left. In order to enable to apply gold studs to the PCB directly its surface has an electroless-nickel-electroless-palladium-immersion-gold coating (ENEPIG). In a flip-chip thermo-compression process the ASIC is then applied to the PCB at a temperature of typically about \SI{200}{^\circ C} and with a bond force of about \SI{5}{\gram} per gold stud, see \autoref{fig:bonding_schematic}, right.
Advantages of gold stud bump bonding are a rather simple and low-cost setup, since the gold studs can be applied directly to aluminium and no UBM is needed, as well as a high deposition rate of 20 bumps per second and a short setup time, which makes is well suited for single die bonding, and thus R\&D projects.
\autoref{fig:bonding_substrates} shows the resulting bumped Timepix and PCB surfaces, and \autoref{fig:bonding_sideview} shows the bonding result as seen in a side view. The two fotos were taken about a centimeter apart and show a constant gap size between PCB and ASIC.

\subsection{Challenges}
\label{sec:hardware_challenges}

\begin{figure}[thp]
  \centering
  \includegraphics[width=\textwidth,height=0.5\textheight,keepaspectratio=true]{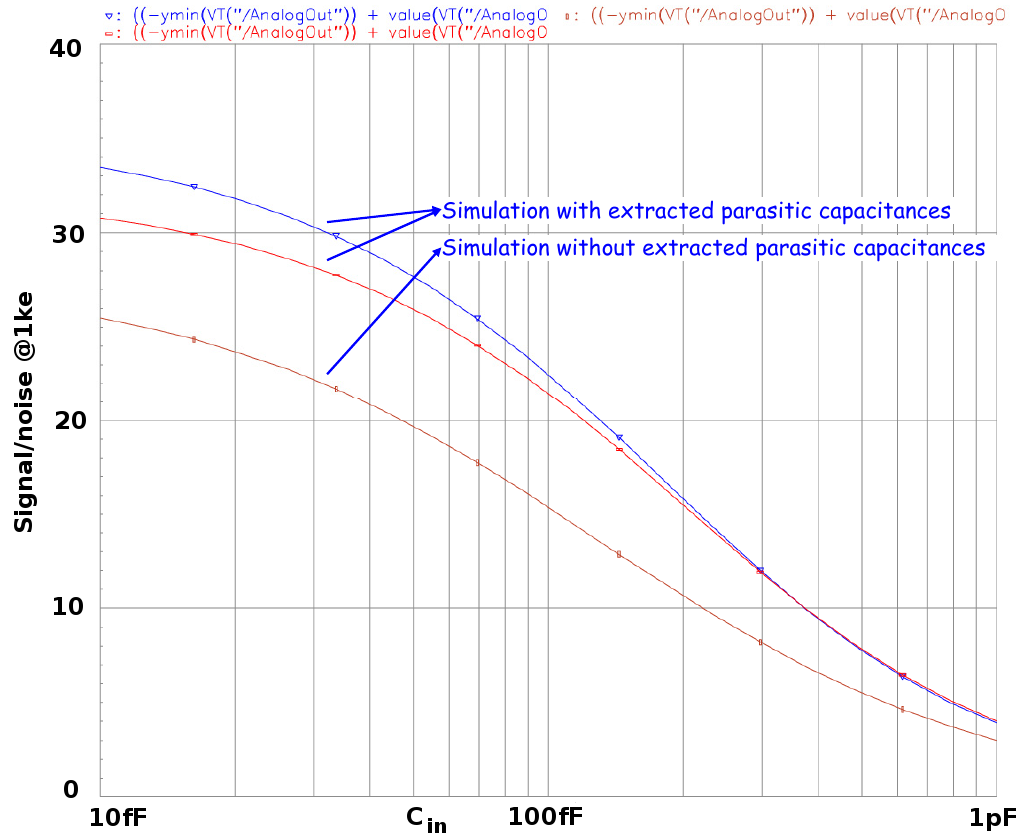}
  \caption{Simulated signal-to-noise ratio of the Timepix ASIC depending on the attached input capacitance, from \cite{Llopart06}.}
  \label{fig:capacitance}
\end{figure}

The discrepancy between the pad pitch of ca. \SI{700}{\micro\meter} and a desired pitch of \SI{300}{\micro\meter} or less results from technical limitations of the routing in the board. The first test board is made from standard FR-4 material, which allows for a minimum feature size of \SI{80}{\micro\meter} and a minimum via size of \SI{300}{\micro\meter}. Further requirements on buried vias and air tightness in the end result in the given limit on the pad pitch. For future boards, other base materials will be taken into account, such as ceramic which has a feature size around \SI{10}{\micro\meter} and a minimum via size of \SI{100}{\micro\meter}. This still does not allow to use the full Timepix ASIC with its \SI{55}{\micro\meter} pitch, but fulfills the cluster finding requirements.
Another challenge is the input capacitance of the pixels: Usually, pixel chips are bonded to sensors with the same pitch and small conductor volume and thus a small capacitance per sensor channel. Therefore, the Timepix ASIC is made for input capacitances below \SI{100}{fF}, compare \autoref{fig:capacitance} \cite{Llopart06}. In case of the ROPPERI system, each connection goes through the board, and the capacitance is dominated by the line length with a value around 1 pF / 25 mm. The pads and the bump bonds add only around \SI{100}{fF} to the input capacitance. For line lengths between pad and pixel in the order of centimeters this clearly reduces the expected signal to noise ratio to a difficult level. This can probably be mitigated by an increased gas amplicifation in the GEM stack, leading to a signal of up to several 10k electrons. In total, a solution to the capacitance challenge seems feasible and will be investigated with the first test board. In particular, the smallest pads are placed directly above the chip to have a capacitance as small as possible, and the medium size pads are designed to have different line lengths to investigate the length dependence of the signal-to-noise ratio.

\subsection{First Measurement}
\label{sec:hardware_first_measurement}

\begin{figure}[thp]
  \centering
  \includegraphics[width=\textwidth,height=0.5\textheight,keepaspectratio=true]{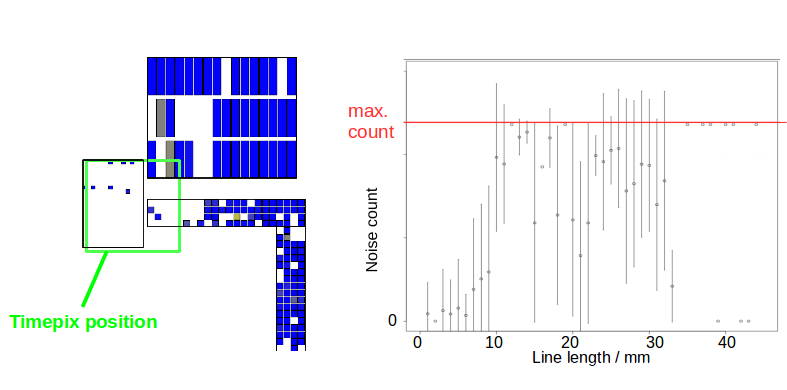}
  \caption{The one reasonable data frame taken before the system broke.}
  \label{fig:one_reasonable_frame}
\end{figure}

After the very first bonding attempt no communication with the Timepix was possible and the ASIC also drew too small currents from the voltage supply, which means there were insufficient communication connections. During the flip-chip process, i.e. while heated to \SI{200}{^\circ C}, a mismatch of roughly \SI{50}{\micro\meter} between the PCB and the chip was observed. This arose from the different coefficients of thermal expansion (CTE) of the materials: While the Si chip has a CTE of 2.5 ppm/K, the PCB material FR-4 lies in the order of 15 ppm/K. A following bonding attempt was consequently done at only \SI{100}{^\circ C}, which worsened the gold-gold-diffusion during the flip-chip process and generated less, yet still substantial mechanical stress. This attempt allowed to take a few very noisy readout frames of the chip and one reasonably good one, before the system broke and data readout was not possible anymore. The very noisy frames were taken immediately after bonding, while the board was still lying on the bonding machine, with the sensitive pads facing the steel surface. Because of the huge capacitance of the steel block, even at very large thresholds all connected channels were noisy - in fact notably, all channels that were supposed to be connected to a sensitive pad were indeed recording noise, and no other channel of the Timepix was.
The one reasonable frame was taken later with the board off the machine. It also contains only noise, but many channels contained a zero signal. The corresponding signal map given in the left part of \autoref{fig:one_reasonable_frame} shows a possible correlation between the noise level and the distance from the Timepix ASIC, which position is depicted in green. On the right, \autoref{fig:one_reasonable_frame} contains the plot of the noise level versus the actual line length as taken from the CAD layout. Indeed there seems to be a correlation: For line lengths below \SI{1}{cm} the noise is inhibited, above that distance the channels are largely at their maximum count of 11810. Also, this frame was taken with a threshold of 380 digital counts, where for (bare) Timepix ASICs the noise can be suppressed by applying a threshold of typically between 300 and 400 counts (above pedestal, one count equals about 25 electrons). This means that the tested setup was at least not dominated by large noise.\\
Since the measurement statistics are small and the errors are large, a solid conclusion is not possible. However, these findings give some confidence regarding the assumptions concerning noise and the manageability of the pixel input capacitance.

\subsection{Next Steps}
\label{sec:hardware_next_steps}

To reduce the thermal stress between PCB and chip new boards with lower CTE were ordered. One consists of VT-901 material, similar to FR-4, layered with copper-invar-copper (CIC) wich has a CTE of 8 ppm/K. The other has as base material N7000-2 HT with a CTE of 12 ppm/K. A larger number of boards were ordered to be able to test various bonding parameters. A second potential source of bad connections is insufficient flatness of the bonding area of the PCB. Therefore, the boards were and are checked for flatness with a laser height measurement. So far, the flatness of the existing boards looks sufficient. The adapted boards are expected in November 2017 and will be repeatedly measured during the setup process. Equipping is done by hand before the bonding to protect the substrates. The first bonding attempt with the new boards is planned soon after their delivery.

\section{Simulation}
\label{sec:simulation}

\begin{figure}[thp]
  \centering
  \includegraphics[width=\textwidth,height=0.35\textheight,keepaspectratio=true]{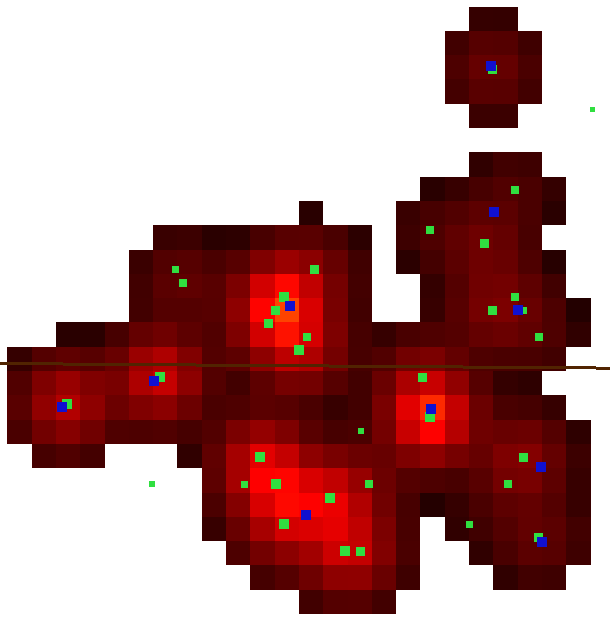}
  \caption{Event display piece of a simulated track (brown line), the drifted electrons (green), the resulting Timepix readout induced by the GEM charge clouds (red, shows granularity) and the hits after reconstruction with the Source Extractor software (blue).}
  \label{fig:reconstruction_example}
\end{figure}

\begin{figure}[thp]
  \centering
  \includegraphics[width=\textwidth,height=0.35\textheight,keepaspectratio=true]{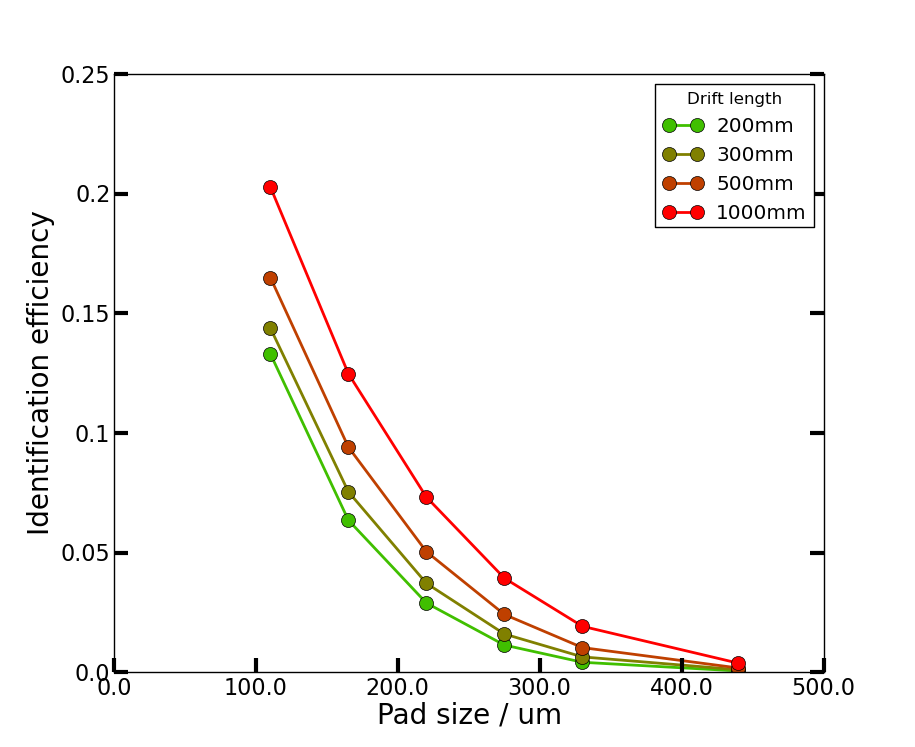}
  \caption{Exact cluster-to-hit-identification efficiency depending on pad size and drift length.}
  \label{fig:cluster_hit_efficiency}
\end{figure}

The prototype system development is accompanied by intense simulation studies to determine the optimal setup of the ROPPERI board, in particular regarding the optimisation of the pad size for cluster identification. A simulation chain has been implemented within the MarlinTPC software framework. It includes the production of MC particles in the detector volume, the drift, amplification and registration of the electrons on the ROPPERI pad plane as well as digitising the signal including noise. Then, a reconstruction algorithm based on the 'Source Extractor' software package \cite{SourceExtractor} follows. This is an external software package originating from astrophysics to scan sky maps for sources and reconstruct them. The algorithm has been adapted to scan for charge clouds in data from the Timepix readout. The necessary data conversion to the .fits-format, the execution of the Source Extractor and the re-import to .slcio were implemented as described in \cite{Deisting12}. The resulting reconstructed hits from the charge clouds are used as input for track reconstruction and dE/dx determination.\\
A major challenge in reconstructing primary ionisation clusters is the high density of electrons from the ionised Argon. \autoref{fig:reconstruction_example} displays different stages of the simulation over a small piece of the readout plane with an exemplary parameter set. The brown line is the original track of a MC particle. The green dots are the positions of the electrons when they arrive at the GEM stack. The brown to red tiles are the resulting charge map of pad raw data induced by the charge clouds, showing also the respective anode granularity. The blue dots are the positions of the reconstructed hits. If the drifted electrons and their respective charge clouds are far enough apart, each one is recontructed as one hit (green dot matches blue dot). Sometimes they are lost in the GEM stack and induce no charge cloud (green dot off the red tiles). Often, the charge clouds overlap and several drifted electrons are reconstruted as one hit (one blue dot in the middle of several green dots). The goal of cluster counting is to identify a group of primary electrons that were generated from a single ionising interaction. Since these groups can split, but also merge, there is no simple way to define a cluster counting efficiency. Instead, using the MC truth information, each drifted electron is assigned to its closest reconstructed hit. One can then calculate the efficiency that this group originates from exactely one primary cluster, and this cluster provides electrons to only exactely one reconstructed hit, called here 'double uniques' or 'exact cluster-to-hit-identification efficiency'. It should be noted, that despite the small number of clusters with many electrons these many electrons tend to contaminate the sorrounding charge clouds of the smaller clusters, which on their own were much more likely to be reconstructed as a double unique, thus reducing the cluster-to-hit-identification efficiency substantially. The cluster-to-hit-identification efficiency depending on the pad size and the drift length are given in \autoref{fig:cluster_hit_efficiency} for a track length of \SI{300}{mm} and a magnetic field of \SI{1}{T}.

%The whole simulation and reconstruction chain has been implemented and first sets of Monte Carlo data have been produced. An example of the Source Extractor output is shown in \autoref{fig:source_extractor}, and in \autoref{fig:point_resolution}, a resulting plot for the point resolution is shown. This work is still ongoing. Especially the reconstruction chain needs to be further tuned to process the data in a more efficient and reliable way.

\begin{figure}[thp]
  \centering
  \includegraphics[width=\textwidth,height=0.35\textheight,keepaspectratio=true]{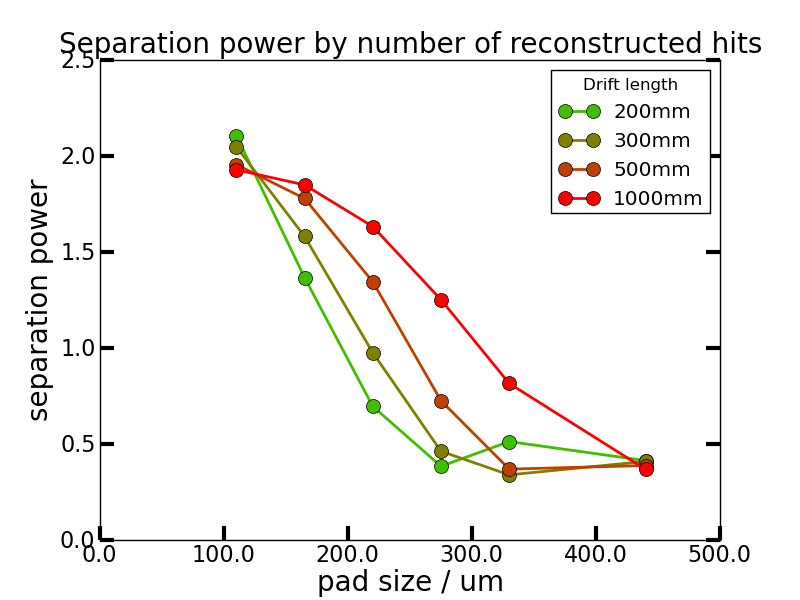}
  \caption{Pion-kaon separation power depending on pad size and drift length, for a particle momentum at the maximum separation.}
  \label{fig:separation_PP}
\end{figure}

\begin{figure}[thp]
  \centering
  \includegraphics[width=\textwidth,height=0.35\textheight,keepaspectratio=true]{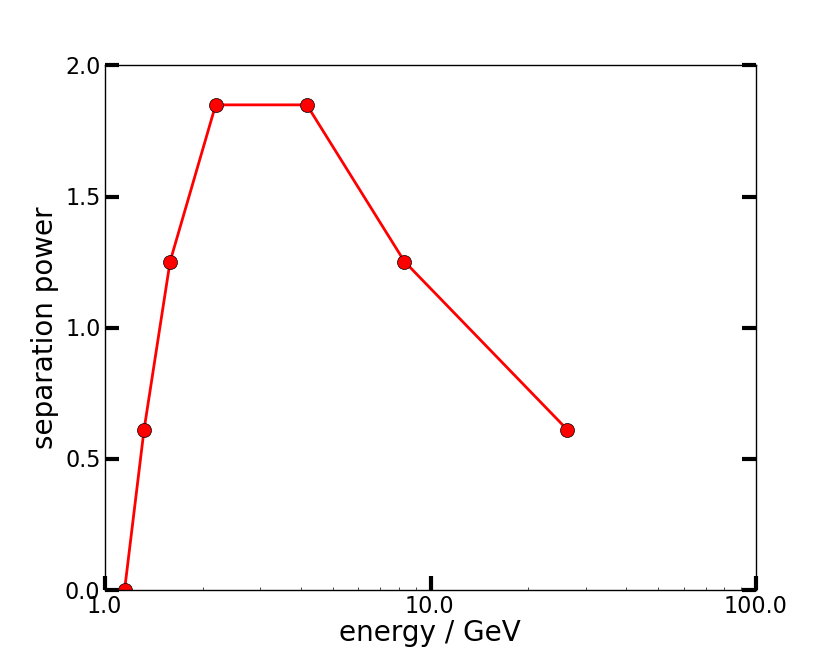}
  \caption{Pion-kaon separation power depending on the particle momentum, for a pad size of \SI{165}{\micro\meter} and a drift length of \SI{1}{\meter}.}
  \label{fig:separation_energy}
\end{figure}

Because of the different definition it is difficult to compare this to the aforementioned cluster counting efficiency in \autoref{fig:piokaonsep}. Instead, one should look at the particle separation power, since this is an unambiguous number and also the goal of a dE/dx measurement. In \autoref{fig:separation_PP}, the pion-kaon separation is shown, for particle momenta at the maximum separation. Conversely, in \autoref{fig:separation_energy}, the separation power is given for different momenta but a fixed pad size and drift length, with its shape corresponding to the right curve in \autoref{fig:piokaonsep}. Assuming the separation power grows with the square root of the track length in the detector, an extrapolation to the envisaged ILD TPC with a track length of \SI{1.2}{\meter} and with a pad size of \SI{165}{\micro\meter} results in a separation power of 3.6, which matches the the 20\% cluster counting efficiency curve and thus represents an improvement over the conventional charge based dE/dx determination.

\section{Conclusion}
\label{sec:conclusion}

The ROPPERI concept allows for a TPC readout with a high level of integration and a high granularity at the same time. The foreseen granularity of a few hundred \si{\micro\meter} is expected to not only provide lower occupancy and better double-track separation, but to also significantly improve the TPC-intrinsic dE/dx capability via cluster counting. At the same time, the advantages of a pad-based readout are preserved, in particular its flexibility.\\
The current hardware development was introduced in detail and first boards have been produced and set up. The very first measurement was shown, including limited insight on the assumptions concerning noise. New boards will be used for prototype tests within the next months, which aim to deliver a proof-of-principle of this technology. In parallel, a software framework has been set up to investigate the future prospects and study more detailed properties using Monte Carlo simulations. First results show promising improvement on particle identification compared to conventional methods.

\section{Acknowledgements}
\label{sec:acknowledgements}

We would like to thank Fabio Colombo from the Karlsruhe Institute for Technology (KIT) for his help regarding the gold stud bonding of the Timepix ASIC.
We thank our DESY colleagues for enduring support, in particular Oliver Schäfer who built the UNIMOCS TPC that will be used for the first tests.

% \begin{figure}[thp]
%   \centering
%   \includegraphics[width=0.95\textwidth,height=0.25\textheight,scale=1,keepaspectratio=true]{ILD.png}
%   ILD.png: 378x340 pixel, 72dpi, 13.34x11.99 cm, bb=0 0 378 340
%   \caption{Overview of the ILD.}
%   \label{fig:ILD}
% \end{figure}

\printbibliography
\end{document}